\documentclass[12pt]{article}
\input{epsf.sty}

\def\fo{\hbox{{1}\kern-.25em\hbox{l}}}

\def\slashchar#1{\setbox0=\hbox{$#1$}           
   \dimen0=\wd0                                 
   \setbox1=\hbox{/} \dimen1=\wd1               
   \ifdim\dimen0>\dimen1                        
      \rlap{\hbox to \dimen0{\hfil/\hfil}}      
      #1                                        
   \else                                        
      \rlap{\hbox to \dimen1{\hfil$#1$\hfil}}   
      /                                         
   \fi}                                         %

\def\hide#1{[hidden stuff]}
\def\bibitemx#1{\bibitem{#1}}

\def\beq{\begin{equation}}
\def\eeq{\end{equation}}
\def\eq{\end{equation}}
\def\to{\rightarrow}

\def\mEt{\mbox{${\hbox{$E$\kern-0.6em\lower-.1ex\hbox{/}}}_T$}\, } 
\def\hinv{h_{\rm inv}}

\def\bsg{\ifmmode B\to X_s\gamma\else $B\to X_s\gamma$\fi}
\def\bsll{\ifmmode B\to X_s\ell^+\ell^-\else $B\to X_s\ell^+\ell^-$\fi}
\def\bstt{\ifmmode B\to X_s\tau^+\tau^-\else $B\to X_s\tau^+\tau^-$\fi}
\def\shat{\ifmmode \hat{s}\else $\hat{s}$\fi}

\newcommand{\newc}{\newcommand}

\newc{\lcal}{\int {\cal L}dt}
 
\newc{\LSP}{{\chi^0_1}}
\newc{\stauR}{{\tilde \tau_R}}
\newc{\stau}{{\tilde \tau_1}}
\newc{\mstop}{m_{\tilde{t}}}
\newc{\mHpm}{m_{H^\pm}}
\newc{\gsim}{\lower.7ex\hbox{$\;\stackrel{\textstyle>}{\sim}\;$}}
\newc{\lsim}{\lower.7ex\hbox{$\;\stackrel{\textstyle<}{\sim}\;$}}
\newc{\ie}{{\it i.e.}}          
\newc{\etal}{{\it et al.}}
\newc{\eg}{{\it e.g.}}          
\newc{\kev}{\hbox{\rm\,keV}}            
\newc{\mev}{\hbox{\rm\,MeV}}            
\newc{\gev}{\hbox{\rm\,GeV}}            
\newc{\tev}{\hbox{\rm\,TeV}}
\newc{\xpb}{\hbox{\rm\, pb}}
\newc{\xfb}{\hbox{\rm\, fb}}

\newc{\mtop}{m_t}
\newc{\mbot}{m_b}
\newc{\mz}{m_Z}
\newc{\mw}{M_W}
\newc{\alphasmz}{\alpha_s(m_Z^2)}
\newc{\swsq}{\sin^2\theta_W}
\newc{\tw}{\tan\theta_W}
\newc{\cw}{\cos\theta_W}
\newc{\sw}{\sin\theta_W}
\newc{\BR}{\hbox{\rm BR}}
\newc{\zbb}{Z\to b\bar}
\newc{\Gb}{\Gamma (Z\to b\bar b)}
\newc{\Gh}{\Gamma (Z\to \hbox{\rm hadrons})}
\newc{\rbsm}{R_b^\hbox{\rm sm}}
\newc{\rbsusy}{R_b^\hbox{\rm susy}}
\newc{\drb}{\delta R_b}

\newc{\sgn}{\mbox{sgn}}
\newc{\tbeta}{\tan\beta}
\newc{\uL}{{\tilde u_L}}
\newc{\uR}{{\tilde u_R}}
\newc{\cL}{{\tilde c_L}}
\newc{\cR}{{\tilde c_R}}
\newc{\tL}{{\tilde t_L}}
\newc{\tR}{{\tilde t_R}}
\newc{\dL}{{\tilde d_L}}
\newc{\dR}{{\tilde d_R}}
\newc{\sL}{{\tilde s_L}}
\newc{\sR}{{\tilde s_R}}
\newc{\bL}{{\tilde b_L}}
\newc{\bR}{{\tilde b_R}}
\newc{\eL}{{\tilde e_L}}
\newc{\eR}{{\tilde e_R}}
\newc{\mhp}{m_{H^\pm}}
\newc{\mhalf}{m_{1/2}}
\newc{\emt}{{e/\mu /\tau}}

\newc{\lR}{\tilde{l}_R}
\newc{\lL}{\tilde{l}_L}
\newc{\nL}{\tilde{\nu}_L}
\newc{\na}{\chi^0_1}
\newc{\nb}{\chi^0_2}
\newc{\nc}{\chi^0_3}
\newc{\nd}{\chi^0_4}
\newc{\ca}{\chi^{\pm}_1}
\newc{\cb}{\chi^{\pm}_2}
\newc{\camp}{\chi^\mp_1}
\newc{\cbmp}{\chi^\mp_1}
\newc{\capos}{\chi^{+}_1}
\newc{\caneg}{\chi^{-}_1}
\newc{\phit}{\phi_t}
\newc{\phib}{\varphi_b}
\newc{\phiew}{\phi_{ew}}
\newc{\htz}{h^0_t}
\newc{\hbz}{h^0_b}
\newc{\hewz}{h^0_{ew}}
\newc{\hsmz}{h^0_{sm}}
\newc{\huz}{h^0_u}
\newc{\hsusyz}{h^0_{susy}}

\newc{\C}{{\cal C}}

\newcommand{\drawsquare}[2]{\hbox{%
\rule{#2pt}{#1pt}\hskip-#2pt
\rule{#1pt}{#2pt}\hskip-#1pt
\rule[#1pt]{#1pt}{#2pt}}\rule[#1pt]{#2pt}{#2pt}\hskip-#2pt
\rule{#2pt}{#1pt}}

\newc{\Dal}{\drawsquare{7}{0.6}}

\def\dofig#1#2{\epsfxsize=#1\centerline{\epsfbox{#2}}}

\def\beq{\begin{equation}}
\def\eeq{\end{equation}}
\def\bea{\begin{eqnarray}}
\def\eea{\end{eqnarray}}
\catcode`@=11
\long\def\@caption#1[#2]#3{\par\addcontentsline{\csname
  ext@#1\endcsname}{#1}{\protect\numberline{\csname
  the#1\endcsname}{\ignorespaces #2}}\begingroup
    \small
    \@parboxrestore
    \@makecaption{\csname fnum@#1\endcsname}{\ignorespaces #3}\par
  \endgroup}
\catcode`@=12

\begin{document}
\begin{titlepage}

\begin{flushright}
UCD-2000-09 \\
LBNL-45356 \\
\end{flushright}






\huge
\bigskip
\bigskip
\begin{center}
{\Large\bf
Pursuing the origin  of
electroweak symmetry \\ breaking: a ``Bayesian Physics'' argument for a \\
$\sqrt{s}\lsim 600\gev$ $e^+e^-$ collider}

\end{center}

\large

\vspace{.15in}
\begin{center}

G.L.~Kane$^a$ and James D. Wells$^b$

\small

\vspace{.1in}
{\it $^{(a)}$Physics Department, 
              University of Michigan, Ann Arbor, MI 48109 \\}
\vspace{0.1cm}
\vspace{0.1cm}
{\it $^{(b)}$Physics Department, 
               University of California, Davis, CA 95616 and \\
Lawrence Berkeley National Laboratory, Berkeley, CA 94720}

\end{center}
 
 
\vspace{0.15in}
 
\begin{abstract}

High-energy data has been accumulating over the last ten years, and it
should not be ignored when making decisions about the future
experimental program.  In particular, we argue that the electroweak
data collected at LEP, SLC and Tevatron indicate a light scalar
particle with mass less than 500 GeV.  This result is based on
considering a wide variety of theories including the Standard Model,
supersymmetry, large extra dimensions, and composite models.  We argue
that a high luminosity, 600 GeV $e^+e^-$ collider would then be the
natural choice to feel confident about finding and studying states
connected to electroweak symmetry breaking.  We also argue from the
data that worrying about resonances at multi-TeV energies as the only
signal for electroweak symmetry breaking is not as important a discovery
issue for the next generation of colliders.  Such concerns should
perhaps be replaced with more relevant discovery issues such as a
Higgs boson that decays invisibly, and ``new physics'' that could
conspire with a heavier Higgs boson to accommodate precision
electroweak data.  An $e^+e^-$ collider with $\sqrt{s}\lsim 600\gev$
is ideally suited to cover these possibilities.

\end{abstract}

\medskip

\begin{flushleft}
hep-ph/0003249 \\
March 2000
\end{flushleft}

\noindent
\begin{center}
{\bf 	Presented at the Berkeley 2000 Workshop \\
Berkeley, California, 29-31 March 2000 }
\end{center}

\end{titlepage}

\baselineskip=18pt
\setcounter{footnote}{1}
\setcounter{page}{2}
\setcounter{figure}{0}
\setcounter{table}{0}


%

\tableofcontents

\section{The many ideas of electroweak symmetry breaking}

The mechanism of electroweak symmetry breaking (EWSB) is still
mysterious.  The ``simplest'' solution is to postulate the existence of
one condensing  $SU(2)$
doublet scalar field that gives masses to the vector bosons
and the fermions.  This  idea is usually spoken of as the
Standard Model explanation for electroweak symmetry breaking.
However, the word ``explanation'' is perhaps too strong.  
The Higgs field provides
no reason why it should have a vacuum expectation value, and it
only exacerbates the hierarchy problem.  Furthermore, the phase
transition associated with the SM Higgs boson solution is not sufficiently
strong first-order to explain the baryon asymmetry of the universe.
These are just three of the reasons why the SM solution is 
unsatisfactory.

A long-standing endeavor in theoretical 
and experimental physics is going beyond
the SM to explain EWSB at a more fundamental level.  The 
would-be explanations
(technicolor, top-quark condensation, supersymmetry, etc.) have invariably
implied new particles and/or interactions with
mass scale near the EWSB scale. 
For example, in strongly-coupled theories such as technicolor or
top-quark condensation the new particles may include
pseudo-Nambu-Goldstone
bosons, and exotic gauge bosons (i.e., new forces) 
and fermions.  In supersymmetry the 
new particles are superpartners and a second
Higgs boson multiplet.  In other words, one should expect additional
particles correlated with a real explanation of EWSB beyond one
physical Higgs boson state.  

The previous two paragraphs could have been written several years
years ago.  What's new today is data.  We sometimes bemoan
the ``lack of data'' in high-energy physics. However, data has been
coming in.  Some theories have died as a result of the data, while
other theories have been emerging as more viable and perhaps 
preferred by the data.  This is what data is supposed to do.  Data
also should help us make decisions about what to look for in the future,
and it should be one of the guiding principles to the future experimental
program. This outlook we call ``Bayesian Physics,'' meaning that 
data and better understanding of theory
is interpreted  to suggest and imply future goals and experiments.  
Emphasis is placed on searching for theories
or classes of theories that have experienced positive success when compared to
data.  The theories do not just survive, they have received positive support
from the data.  Supersymmetry is in this class, we believe.

A competitor
philosophy is what we call ``Nonjudgmental Physics.''  In this philosophy,
any physics idea (complete or incomplete)
that still is technically not dramatically excluded
by the data is equally likely.
This outlook dictates, for example, that we view 
a light-Higgs predicting theory
(supported by recent data) as equal in stature to a
strong EWSB sector theory (not supported by recent data, and
probably but not necessarily ruled out) 
when thinking about the requirements of future
experiments.  This philosophy is clearly the superior philosophy in
the limit of infinite amount of time, money, and people to do experiments.
We believe that the ``Bayesian Physics'' outlook can help set priorities
when resources are limited, and may be essential to obtain
new scientifically useful facilities.

\section{Indications of a light Higgs boson}

There are several important inputs from data that can help us
decide what theories are more probable.  Gauge coupling
unification, for example,
can be interpreted as a great success for supersymmetric theories.
This alone may be powerful enough for some to consider only theories
consistent with supersymmetry gauge coupling unification.  Although
we interpret gauge coupling unification as a powerful message
that supersymmetry is part of nature, we will not dwell on this subject
here.  Instead, we wish to interpret the precision EW data in a wider
class of theories.  

We argue in this section that all theories that support a 
Higgs boson (composite or fundamental)
with mass significantly less than about $500\gev$ should be given special
status over all other candidate theories of nature.  Therefore,
when making decisions about new collider facilities, it makes more
sense to discuss and analyze all the vagaries of lower mass Higgs
bosons than it does to compare a simple SM Higgs signature with
signatures of a strongly coupled EWSB sector at very high energies.
We will motivate this viewpoint in the next few paragraphs, and in the
next sections we will outline some of  the relevant discovery issues.

Since a strongly-coupled EWSB sector contributes to precision electroweak
observables like a $\sim{\rm TeV}$ 
Higgs boson~\cite{Chanowitz:1998wi}, it appears
reasonable to declare these theories as less likely than other theories
with fundamental or composite light Higgs bosons.  Of course, it might be
possible to construct a fully consistent theory that combines strongly
coupled EWSB with other new physics that conspires to describe
electroweak precision data.  However, the new physics that 
accompanies such theories usually {\it worsens} the predictions
for precision observables.  For example, large positive
contributions to the $S$ parameter in technicolor theories combined
with the effective high mass of a ``Higgs boson'' are incompatible with
the data.  In our view, strongly coupled EWSB, although not necessarily
ruled out, has become a less-motivated concern given the current data.
We will therefore not discuss further this possibility, and rather 
focus on the more data-motivated scenarios of light fundamental or
composite Higgs bosons.

First, the EW precision data from LEP/SLD over the last ten years 
and direct searches for Higgs bosons at LEP2 
indicate that a SM-like Higgs boson with mass between
$110\gev$ and $215\gev$ is a good fit to the 
data (95\% C.L.)~\cite{lep ewwg}.
This is our main input to the discussion.  This lends support to
any theory that predicts a SM-like Higgs boson with mass less
than $215\gev$ and above $110\gev$.  Supersymmetry is one 
such theory; it certainly predicts $m_h<215\gev$, and a part
of parameter space allows $m_h>110\gev$. Actually, the lightest
physical Higgs boson of the general supersymmetry case can be as
light as 85 GeV and consistent with LEP data.  
Also, there are no internal
inconsistencies with the SM up to very high scales if its mass 
is in this range.  In fact, if such a light SM-like Higgs boson
can be taken at face
value it may be possible to detect a Higgs boson at the Tevatron
even before LHC and Tevatron~\cite{fermilab workshop}.

A zeroth-order conclusion of
a ``Bayesian Physics'' view of the data supporting a light Higgs boson
is to build a collider
that can find and study a Higgs boson with mass less than $215\gev$.
However, this may be a bit naive since the precision data is measuring
virtual effects that may include cancellations and combinations of many
different kinds of states that in the end imitate the effects of
a light Higgs boson.  In the next section we will discuss these possible
conspiracies and show that they imply that the Higgs boson could be as
heavy as $500\gev$, but not more.

\section{Cancellation conspiracies for a heavy Higgs boson}

The light Higgs boson requirements of precision EW data based on
a SM analysis can be imitated by other effects.    
Several groups~\cite{Hall:1999fe}--\cite{Marciano:2000yj}
have studied the possibility of raising the Higgs boson mass substantially
above $215\gev$ and using other states or operators to cancel the effects
of this heavier Higgs mass in the radiative corrections, thereby allowing
agreement with electroweak symmetry breaking.

One example of this type of conspiracy is found in ref.~\cite{Rizzo:2000br}.
The theory is the SM in extra dimensions, where
the fermions live on a 3+1 dimensional wall, and the gauge bosons live
in a higher dimensional space~\cite{Antoniadis:1990ew,Pomarol:1998sd}.   
If the Higgs boson lives on the 3+1
dimensional wall with the SM fermions, it can mediate a mass mixing between
the ordinary $W,Z,\gamma$ gauge bosons and their Kaluza-Klein excitations.
This mass mixing then leads to shifts in the EW precision observables
that can mimic the effects of a light Higgs boson~\cite{Rizzo:2000br}.  
That is, a heavy Higgs
boson plus gauge boson mode mixing leads to predictions for $Z$ pole
observables very similar to that of a light Higgs boson.

Any prediction for an
observable ${\cal O}_i$ ($\Gamma_Z$, $\sin^2\theta^{\rm eff}_W$, etc.) can be
expanded approximately as
\beq
\label{observable}
{\cal O}_i={\cal O}_i^{\rm SM}+ a_i \log m_h/m_Z + b_i V.
\eeq
${\cal O}_i^{\rm SM}$ is defined to be the best fit 
value of the observable assuming
the SM and $m_h=m_Z$. We choose $m_h=m_Z$ arbitrarily for this expansion,
but it is also convenient since it is approximately
the value of $m_h$ at the global minimum of the fitting $\chi^2$,
\beq
\chi^2 = \sum_i \frac{({\cal O}_i-{\cal O}_i^{\rm expt})^2}
            {(\Delta {\cal O}_i^{\rm expt})^2}.
\eeq
$V$ represents the
effects on the observable from 
Kaluza-Klein excitations of the gauge bosons, and is defined as
\beq
V\equiv 2 \sum_{\vec n} \frac{g^2_{\vec n}}{g^2}\frac{m_W^2}{\vec n^2 M_c^2},
\eeq
where $M_c=R^{-1}$ is the compactification scale of the extra
spatial dimension(s). If there is only one extra dimension then
\beq
V=\frac{\pi^2}{3}\frac{m_W^2}{M_c^2}.
\eeq

The measurements of the observables ${\cal O}_i^{\rm expt}$
are in good agreement with the SM prediction ${\cal O}_i^{\rm SM}$
as long as $110\gev \lsim m_h\gsim 215\gev$.  
If $b_i V=0$ (decoupled effects of KK excitations)
then ${\cal O}_i$ is merely the prediction of
the SM for some $m_h$.  If $m_h$ gets too large, ${\cal O}_i$
gets further away from the best-fit value of $m_h\simeq m_Z$ and the prediction
does not explain the data.  However, if $b_i V\neq 0$, it is possible
to have a cancellation between the $\log m_h/m_Z$ and $V$ terms 
in Eq.~(\ref{observable}) even for $m_h\gg m_Z$.  This would constitute
a ``conspiracy'' of cancellations to allow a large Higgs mass.

The trouble with conspiracies is that the cancellation must occur
for {\it every} well-measured observable.  Although cancellations can
be arranged in some observables to maintain the light-Higgs SM prediction
for larger Higgs mass and larger $V$, the cancellation cannot be maintained
for all observables.  The $\chi^2$ 
function may stay under control for somewhat larger Higgs masses due to
this cancellation adjustment among the most precisely measured observables,
but at some point the less-precisely measured observables will deviate too
far from the experimental measurements and cause the $\chi^2$ to rise
unacceptably high.

As an example of the general statements of the last few paragraphs, we
show  how the Higgs mass limit in extra dimensions can be increased
above the SM limit but not to arbitrarily high values.  The most precisely
measured observable relevant to Higgs boson physics is 
$\sin^2\theta_W^{\rm eff}$.  It can be expanded as
\beq
\sin^2\theta_W^{\rm eff}=\sin^2\theta_W^{\rm eff,SM}
  +0.00053\log m_h/m_Z - 0.44 V.
\eeq
Again, $\sin^2\theta_W^{\rm eff,SM}$ is the SM best-fit value for
$m_h=m_Z$, which is in good agreement with the experimental measurement.
To maintain this good agreement for higher Higgs mass, $V$ must
satisfy
\beq
\label{relate V and mh}
V= 1.2\times 10^{-3}\log m_h/m_Z .
\eeq
For example, if $m_h=500\gev$ then $V=2.0\times 10^{-3}$, which corresponds
to a compactification scale of $3.3\tev$ for one extra dimension. The
compactification scale is also the mass of the first Kaluza-Klein excitations 
of the gauge bosons.  Indeed the analysis of ref.~\cite{Rizzo:2000br}
demonstrates this general relationship between $m_h$ and $V$ as
derived in Eq.~(\ref{relate V and mh}).  However, $m_h=500\gev$ is
right at the edge of a tolerable {\it total} $\chi^2$ for all
precision observables.  That is, the cancellation between large Higgs mass
effects and Kaluza-Klein gauge boson effects is only partially working
for other observables.  

Another important observable is
$m_W$.  The theoretical prediction can be expanded similarly
as we did for $\sin^2\theta_W^{\rm eff}$,
\beq
m_W = m_W^{\rm SM} - (0.07\gev)\, \log m_h/m_Z + (34\gev)\, V.
\eeq
Plugging in $m_h=500\gev$ and $V=0.002$ we get
\beq
m_W=m_W^{\rm SM} - 0.12\gev + 0.07\gev = m_W^{\rm SM} - 0.05\gev.
\eeq
There is still a cancellation effect between heavy Higgs and light
compactification scale in $m_W$, but it is not complete.
The parameters have subtracted $50\mev$ from the SM light-Higgs prediction.
The measured value~\cite{lep ewwg} and the 
SM best-fit prediction for $m_W$ with $m_h=m_Z$ are
\beq
m_W^{\rm expt}=80.419\pm 0.038\gev ~~~~{\rm and}~~~~
             m_W^{\rm SM}=80.395\gev .
\eeq
Subtracting $50\mev$ from $m_W$ is clearly a prediction
that does not match the measurement very well, and the large Higgs mass
starts to run into trouble in the fit to EW parameters.

The above example is made more rigorous by doing a complete $\chi^2$ analysis
of the data using the parameters of the extra dimensional 
theory~\cite{Rizzo:2000br}.  The result is that Higgs boson masses can
be extended beyond the SM mass limit, but only up to $500\gev$ at the
95\% C.L.  For $m_h>500\gev$ the theory is not a good match to the data.

We believe that the extra-dimensional example illustrates a 
general lesson.
That is, there are too many observables precisely measured to expect a 
global conspiracy cancellation between heavy Higgs effects (fundamental
or composite) and other physics contributions.  It may be possible to have
a collection  of higher-dimensional operators conspire to allow a larger
Higgs mass~\cite{Hall:1999fe,Bagger:1999te}, but these examples are
not real theories, and there appears to be no motivation for choosing
the considered effective Lagrangian other than to construct this
cancellation.
Furthermore, one can argue generally that these conspiracies of operator
coefficients are
unlikely~\cite{Barbieri:1999tm,Kolda:2000wi,Marciano:2000yj}.  

Other specific example theories that demonstrate 
cancellation of the effects of a larger Higgs mass
have been proposed in the literature~\cite{PeskinWells}.  One such
theory
is the see-saw top-quark condensate 
model~\cite{Dobrescu:1998nm,Cheng:1999bg,Chivukula:1999wd}, 
and the cancellation occurs
between a heavy composite Higgs boson and the virtual effects of a massive
quark mixing with the top quark~\cite{Collins:2000rz}.  
However, as shown in~\cite{Chivukula:2000px}, Higgs
masses above $500\gev$ are not allowed and still maintain theoretical
consistency.  Furthermore, ref.~\cite{Chivukula:2000px} has demonstrated
an approximate $450\gev$ mass limit on conspiring Higgs composite models.
It is interesting that in all the more detailed, independent 
studies of conspiring
theories such as the extra dimensional theory~\cite{Rizzo:2000br},
and the composite theories~\cite{Chivukula:2000px}, the mass limit
of $m_h<500\gev$ survives.  And, of course, the minimal supersymmetric
standard model automatically predicts a light SM Higgs boson.  In all
these cases, the indications from the data and a wide range of theory point
to a Higgs mass below $500\gev$.  We think this is the goal to shoot
for in a high energy collider program.

\section{Resolving the ``new physics''}

We have argued in the previous section that it is likely that a light Higgs
boson exists and accounts for the precision EW data taken at the Z-pole
and elsewhere. If its production and decay are close to that of the SM
Higgs boson, both the NLC and the LHC could discover it.  Nevertheless,
the NLC would usher in an extraordinary era of precision Higgs boson physics,
that would be useful in studying the dynamics
and structure of EWSB symmetry breaking.  To some, this is powerful enough
reason to support an NLC program.

However, we would like to point out that there are important {\it discovery}
issues surrounding a light Higgs boson.  One issue that we will address
in the next section is an invisibly decaying Higgs 
boson~\cite{theory invisible higgs}.  This possibility
is certainly not a ridiculous theoretical musing, and we as a community
should make sure that it is covered experimentally.   The other discovery
issue we would like to discuss is the ``new physics'' that conspires to
allow a heavier Higgs boson satisfy the precision EW data.  What is the
most effective way to discover the nature of such new physics?

Above we gave two concrete examples of new physics that could conspire to allow
a Higgs boson up to 500 GeV.  One example is a top seesaw
model with one $Y=4/3$ fermion in addition to the SM fermions, 
and the other example is large extra dimensions for the gauge fields
with the Higgs boson living on a $3+1$ dimensional wall with the fermions.
We will consider each in turn.

First, we consider the top-quark seesaw model with one extra
fermion with hypercharge $4/3$ as analyzed in ref.~\cite{Collins:2000rz}.  
This fermion participates
in a condensate seesaw with another quark to produce one light eigenvalue,
the top quark $t$ with mass $m_t\simeq 175\gev$, and 
one heavy eigenvalue, $\chi$.  The effects of $\chi$ on precision
EW analysis is such that it could conspire with a heavy composite
Higgs boson to mimic the effects of a light Higgs boson.  
Actually, this statement is not precisely correct since varying the Higgs boson
mass maps out a {\it different} path in the $S$-$T$ plane, for example, than
the path generated by varying the $\chi$ mass. $S$ and 
$T$ are defined~\cite{Peskin:1990zt} by
\bea
\frac{\Pi_{ZZ} (m^2_Z)-\Pi_{ZZ}(0)}{m^2_Z} & = & 
     \frac{\alpha (m_Z)S}{\sin^2 2\theta_W (m_Z)} \\
\frac{\Pi_{WW}(0)}{m^2_W}-\frac{\Pi_{ZZ}(0)}{m^2_Z} & = &
  \alpha(m_Z) T
\eea
where all parameters are in the MS-bar scheme.  

The SM prediction for
$S$ and $T$ depends on $m_t$ and $m_h$ as well as other parameters
of the theory.
With the reference values $m_t^{\rm ref}=175\gev$ and
$m_h^{\rm ref}=500\gev$ for the SM parameters we can calculate
the prediction of $S$ and $T$: $S_{\rm SM}^{\rm ref}$
and $T_{\rm SM}^{\rm ref}$.  The experimental best fits to $S$ and
$T$ are~\cite{Bagger:1999te}
\bea
S_{\rm expt} & = & S-S^{\rm ref}_{\rm SM} = -0.13\pm 0.10 \\
T_{\rm expt} & = & T-T^{\rm ref}_{\rm SM} = 0.13 \pm 0.11.
\eea
The $68\%$ and $95\%$ C.L. contours for this fit are given in Fig.~1.
We have also put X marks on the plot for SM prediction with 
$m_h=100\gev, 200\gev, \ldots , 1000\gev$ going from left to right.
As we can see from the plot, the 95\% C.L. bound on the Higgs boson
in the SM is between $200\gev$ and $300\gev$, consistent with the
value $229\gev$ obtained in ref.~\cite{Bagger:1999te}.

The phenomenology of the one-doublet top seesaw model is very similar to the
SM with one extra, massive quark $\chi$. If light, this quark contributes
substantially to $T$ but very  little to $S$.  Nevertheless, one could
imagine a  large Higgs mass conspiring with a smaller $\chi$ mass to
generate a good fit to the data. We demonstrate an example of this by 
supposing that the Higgs boson has mass of $m_h=500\gev$ and the
$\chi$ has mass of about $5\tev$.  Then, the  shift in $\Delta T_\chi$
can put the theory prediction well within the 95\% C.L. contours
of precision EW fits.  This is the origin of the claim~\cite{Collins:2000rz}
that Higgs boson masses above $300\gev$ are not in conflict with the
data as long $5\tev\lsim m_\chi\lsim 7\tev$.

\begin{figure}[t]
\dofig{6in}{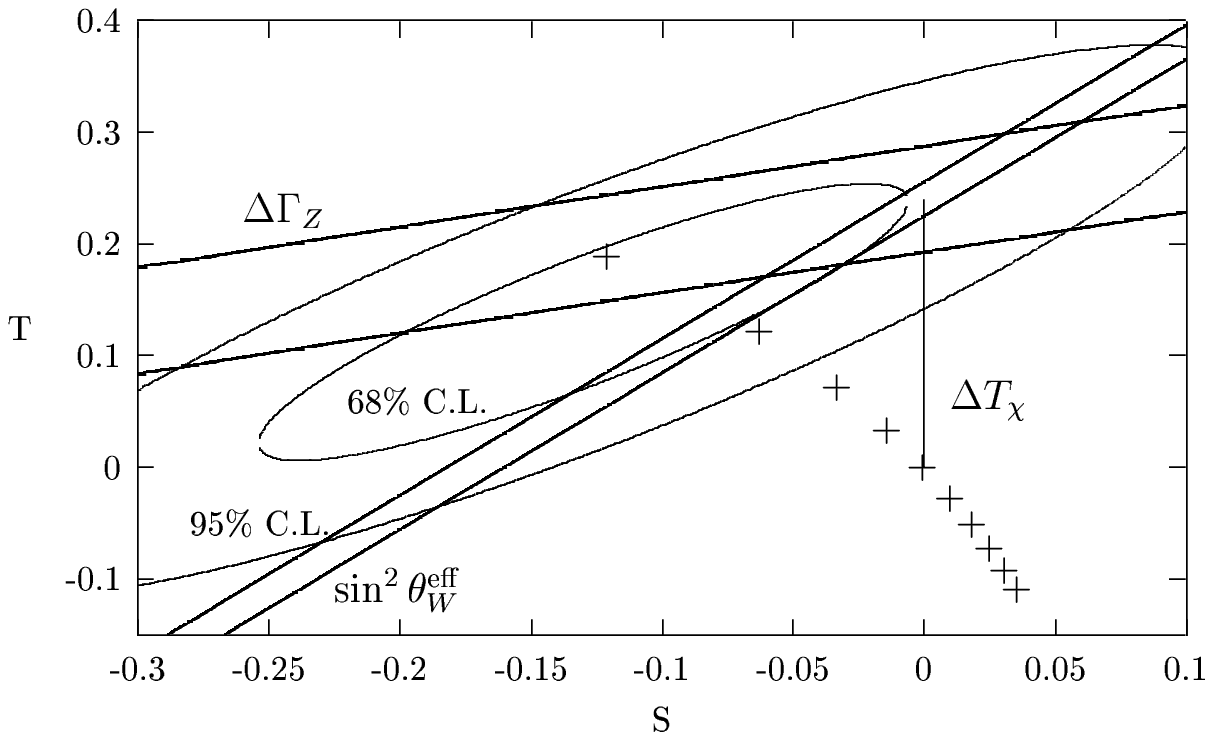}
\caption{
Best fit values to $S$ and $T$ relative to the SM reference
point with $m_t=175\gev$ and $m_h=500\gev$.  The larger circles
represent current constraints as calculated by~\cite{Bagger:1999te}.
The $X$ marks on the graph refer to Higgs boson masses
$100\gev, 200\gev, \ldots, 1000\gev$ from left to right.  The vertical
line labelled $\Delta T_\chi$ represents a
$m_\chi \simeq 5\tev$ new quark conspiring with $m_h=500\gev$ for
an acceptable fit to the current data.
The steeper narrow strip illustrates the 95\% C.L. band allowed
by a measurement of $\Delta \sin^2\theta = 0.00004$ (95\% C.L.),  and the
less steep strip illustrates  the 95\% C.L. band allowed
by a measurement of $\Delta \Gamma_Z= 1\mev$ (95\% C.L.).  These values
may be obtained at the NLC if more than about $50\xfb^{-1}$ is obtained
on the $Z$ pole.
\label{stplane}}
\end{figure}

\begin{figure}[t]
\dofig{6in}{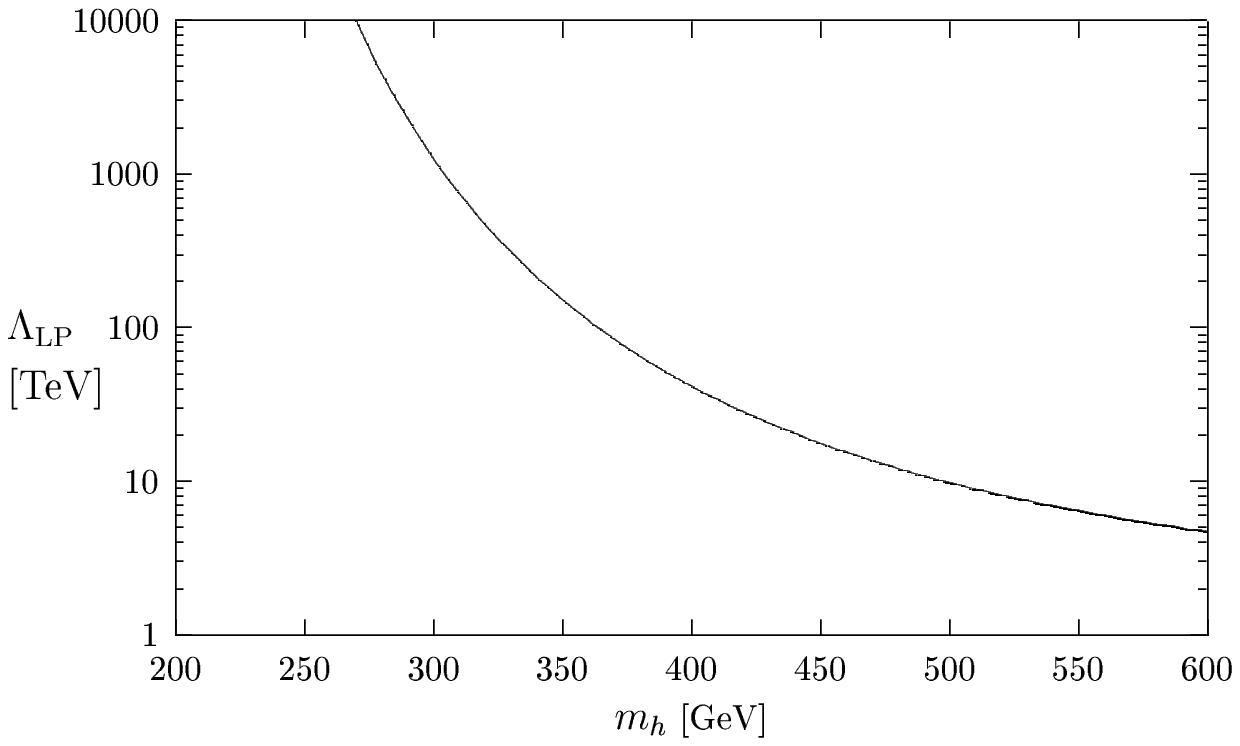}
\caption{
$\Lambda_{\rm LP}$ is the scale at which the Higgs boson self-coupling
reaches its Landau pole (blows up).  It is correlated with higher
dimensional operators that can alter the predictions for precision
electroweak observables.  As $m_h$ goes higher, $\Lambda_{\rm LP}$ gets
lower and runs an increasing risk of disrupting good agreement with
the data.  For $m_h\geq 500\gev$, $\Lambda_{\rm LP}\lsim {\rm several}\tev$
which generally implies too large corrections to EW observables.
\label{landau pole}}
\end{figure}

There are several lessons to learn from this example in our opinion.
First, the Higgs mass in this theory is not expected to be above
$500\gev$~\cite{Chivukula:2000px} in any event. We can understand this
result by correlating the Higgs mass with the Landau pole
scale $\Lambda_{\rm LP}$ where the Higgs self-coupling blows up.
We plot $\Lambda_{\rm LP}$ vs.\ $m_h$ in Fig.~\ref{landau pole}. 
The scale $\Lambda_{\rm LP}$ directly correlates with other parameters
in the top seesaw model, most notably the extra fermion mass which
must be below $\Lambda_{\rm LP}$ in order for condensation
to occur.  Therefore,
knowing $\Lambda_{\rm LP}$ enables us to determine the effects of
new physics on precision electroweak observables.  As shown 
in~\cite{Collins:2000rz,Chivukula:2000px}, the custodial symmetry violations
associated with the new fermion mixing with the top quark induce
a large contribution to the $T$ parameter proportional to $m_Z^2/m_\chi^2$.
The coefficient of this proportionality has been estimated, and a
conservative conclusion is that no set of parameters with Higgs mass
greater than $500\gev$ will allow a good fit to precision electroweak data.
In other words, as $m_h$ gets higher $\Lambda_{\rm LP}$ gets lower,
and as $\Lambda_{\rm LP}$ gets lower $m_\chi$ gets lower and causes
$T$ to be much too large to accomodate the precision electroweak data.
We also note that
from the discussion of limits on the coefficients  
of higher-dimensional operators (usually $\sim 1/(8\tev)^2$ for dimension
six operators~\cite{Barbieri:1999tm}), 
it appears unlikely that conspiracies will be effective
for any theory with a Higgs mass greater than $500\gev$.

The $500\gev$ limit we have discussed for the top seesaw model is
a specific case of the more general Chivukula-Simmons 
bound~\cite{Chivukula:1996sn}.  This bound states that when you
correlate minimal custodial symmetry breaking requirements with the
Higgs Landau pole scale, the experimentally measured
bound on the $T$ parameter imply
stronger bounds on the Higgs boson than mere triviality. Note, this bound
is not purely theoretical and requires the important input
of precision electroweak data.  

Nevertheless, a larger Higgs mass of up to $500\gev$ can conspire
to bring the prediction back into the
allowed region in the $S$-$T$ plane.  However, this is not a
{\it cancellation} of the effects of a large Higgs mass.  The pathway
made in the $S$-$T$ plane by a variable Higgs mass is significantly
different than that made by varying other parts of the theory, in
this case the $\chi$ mass.  Better precision low-energy measurements would be
able to distinguish the two theories.  In the plot we anticipate a
a reduction of errors on $\sin^2\theta^{\rm eff}_W$ and
$\Gamma_Z$ by running on the $Z$ pole with over $50\xfb^{-1}$ of
integrated luminosity at the NLC.  Using ref.~\cite{Hawkings:1999ac} 
as our guide,
we think it might not be unreasonable to obtain
$\Delta \sin^2\theta^{\rm eff}_W=0.00002$ and
$\Delta\Gamma_Z=1\mev$ at the 95\% C.L.  The first estimate is well-within
the anticipation of~\cite{Hawkings:1999ac}; 
however, the second number is a factor
of 4 better than cited by~\cite{Hawkings:1999ac}.  
The error on $\sin^2\theta^{\rm eff}_W$ may be dominated by
uncertainty in $\alpha_{\rm QED}(m_Z)$.  It is best reduced by
doing precise scans of $e^+e^-\to {\rm hadrons}$ at low energies,
and a discussion of the experimental plans, expectations, and hopes
can be found in ref.~\cite{Zhao:2000ev}. 
The error on $m_Z$
is dominated by the uncertainty of $\sqrt{s}$ and $1\mev$ at the
95\% C.L. may not be doable.  Nevertheless, even with just the
$\sin^2\theta_W^{\rm eff}$ measurement, one should be able to
test consistency of a Higgs mass measurement with the predictions
for $S$ and $T$, and find a descrepancy,  implying new states. 
The more precise
measurement of $\Delta\Gamma_Z$ would help even more dramatically pin
down the type of new physics that is compensating for the heavier
Higgs boson. We also see from the graph, that the heavier the
Higgs boson mass, the easier it should be
to unravel any conspiracy with more $Z$-pole data.
For this reason, we encourage additional study of the
$Z$ pole precision EW measurement capabilities at the NLC.

We also learn that 
although conspiracies do correlate with light ``new physics'' this
does not guarantee that the new states will be directly produced and
discovered at the next generation colliders. 
In the top seesaw model that we are considering
now, neither the NLC nor the LHC will be able to directly observe $5-7\tev$ 
quarks. Only high-luminosity precision $Z$-pole measurements would really
be able to see the evidence for new physics by constraining, for example,
$S$ and $T$ to be off the Higgs boson path.  With a precise measurement
of the Higgs boson mass, the trajectory of the new physics in
the $S$ and $T$ plane could be determined.  We speak in terms of $S$ and $T$
here because it is valid and useful in this example, but in a more
general approach one could analyze the multidimensional space of
observables with all the self-energy and vertex corrections included.

The second example is large extra dimensions.  We stated in the previous
section that light Kaluza-Klein modes of the gauge bosons could conspire
with a large Higgs mass to satisfy EW precision data.  The global
$\chi^2$ implied that $m_h<500\gev$ is required.  If a Higgs mass were
discovered with mass somewhere between $400\gev$ and 
$500\gev$, then one would expect in this scenario to
find gauge boson KK states starting somewhere 
between $3.3\tev$ and $6.6\tev$.  
This is clearly out of the reach
for direct detection at the NLC, but the LHC will be able to see
KK excitations up to at least $5.9\tev$ with $100\xfb^{-1}$, 
covering a significant portion
of the parameter space.  On the surface this appears to be bad news
for the NLC and good news for the LHC.  However, the precision measurement
capabilities of NLC at high energies
allows one to be sensitive to virtual, tree-level
exchanges of KK states in $e^+e^-\to Z^{(n)}/\gamma^{(n)}\to f\bar f$.
The sensitivity to KK states using all the observables at one's disposal
at the NLC is extraordinary.  A $600\gev$ NLC with $50\xfb^{-1}$
can see the effects of KK excitations well above $10\tev$~\cite{Rizzo:2000br}. 
Furthermore,
one can show that the LHC will have an extremely difficult time resolving
the degenerate KK modes from an ``ordinary'' $Z'$ gauge boson.  However,
the NLC can resolve the difference~\cite{Rizzo:2000en}.  

Also, additional ``new physics'' discoveries might be  possible 
only through production and subsequent decay of
Higgs bosons.  This could be the case if a Higgs boson decays into
neutrinos or graviscalars in extra dimensions.
Probably any decay mode of even the ``heavier'' Higgs bosons of conspiracy
theories would still allow discovery.    They are usually produced the
traditional ways, in $Z/W+h$ associated production, $gg\to h$,
and $WW\to h$.  Below the
top threshold they decay mainly to $WW$ and $ZZ$ and invisibly, and above the
top threshold mainly to $t\bar t$ and possibly invisibly.  At LHC they will
be produced, but detecting them may be very difficult, particularly if
the mass is above
the top threshold.  The invisibly decaying Higgs boson will be discussed
in somewhat more detail in the next section. 
NLC has a significant advantage here.

In short, both theories discussed have been touted as explicit realizations
of a conspiracy to accomodate a heavier Higgs boson in precision EW data.
However, both theories upon close examination prefer the Higgs boson
not be heavier than $500\gev$, kinematically accessible to 
a $600\gev$ NLC.  Furthermore, and perhaps most importantly, any new physics
that contributes to the conspiracy is more likely to be discernible at the
NLC than the LHC.

We have said very little about supersymmetry in this paper, even though
we have more confidence in its relevance to nature than the other theories
discussed.  Supersymmetry has been well-established to predict a light-Higgs
boson in the spectrum, easily accessible at the NLC.  The ``new physics'' of
supersymmetry are the superpartners.  Unlike many other ideas,
supersymmetry is a 
well-defined, perturbative gauge theory, and it is possible to rationally 
study issues such as finetuning of electroweak symmetry 
breaking~\cite{finetuning}.  
Numerous studies are in agreement  that at least some superpartners should
be less than a few hundred GeV. Furthermore, if the lightest supersymmetric
partner is stable then cosmological constraints generally, but not always, 
imply
upper bounds on superpartners accessible at the 
NLC~\cite{Wells:1998ci,ellis separate}.  In our view it is
rather obvious that supersymmetry enthusiasts would support
the NLC, if for no other reason than to study the Higgs 
boson properties carefully.  Our discussion above is meant for those who
worry about a broader perspective.


\section{Invisibly decaying Higgs boson is more pressing concern}


In discussing the NLC capabilities of discovering
and studying EWSB, one is often led to comparing with the LHC.
The discussions usually begin with noting
that kinematically accessible
states will be studied very effectively at the NLC and decay branching
fractions and production cross-sections will be measured to impressive
accuracy.  However, at this stage most of us are concerned with discovery,
and so it is frequently brought up that the NLC will have 
a hard time with strongly coupled EWSB theories, where resonances
at perhaps several TeV would be the only experimental indication
that the $W_LW_L$ scattering cross-section is being unitarized.
This has traditionally been
implicitly thought of as the best example of a ``problematic
non-SM-like EWSB signal that must be covered''.  The NLC typically
struggles in this analysis.

However, we feel that the ``metric'' on all possible beyond-the-SM theories
is grossly distorted by contrasting different collider's ability to
discover and study either a SM-like light Higgs boson or difficult
multi-TeV resonance signals. There are many
more {\it discovery} issues than strongly coupled EWSB sectors.  And,
from our discussion in the introduction and the previous section, we
believe that the relevance of multi-TeV resonance signals has diminished
dramatically given the data collected on the $Z$ pole over the last
ten years.

We would therefore like to emphasize other potential discovery issues
for Higgs bosons.  There are many possible discovery challenges for
even light Higgs boson(s).
Perhaps the most important ``problematic
non-SM-like EWSB signal that must be covered'' is an invisibly decaying
Higgs boson. There are several well-motivated theoretical 
reasons~\cite{theory invisible higgs} why
a Higgs boson may preferentially decay into invisible, non-interacting
states.  These accessible decay modes certainly do not 
need to affect the Higgs couplings
to gauge bosons or SM fermions, and so precision EW observable analyses
would follow through just as for the SM Higgs boson.  However, the decays
will cause problems for the detectability of the Higgs boson itself.
If we want to discover and study the Higgs bosons we should analyze carefully
the prospects for discovering this rather difficult possibility.
Similarly, the Higgs may decay invisibly only part of the time, 
which actually could make discovery more difficult at both NLC and LHC.

Furthermore, within the context of supersymmetry, there are many
additional ways that Higgs boson {\it detectability} could be a major
challenge to high-energy colliders.  For example, a complex Higgs
sector, with many physical light Higgs states may escape all detection
at the LHC, and also be a challenge at the NLC.  However, with sufficient
luminosity, a $500\gev$ NLC should be guaranteed to see a Higgs boson
signal~\cite{Espinosa:1999xj}.

Returning to the single invisibly decaying Higgs case,
there have been several analyses evaluating discovery possibilities
at hadron colliders and lepton colliders.  First, LEPII collaborations
have published searches~\cite{expt invisible higgs} 
for such states and generally
get limits near $m_{\hinv}<99\gev$ 
assuming SM strength coupling to the $Z$ boson and $100\%$ branching
fraction into invisible final states.  
Future runs at LEPII will not go much beyond
this number.  Nevertheless, the limit is very close to the kinematic
edge $\sqrt{s}-m_Z$ from $e^+e^-\to \hinv Z$.  The Tevatron
presently has no meaningful
limits on the invisibly decaying Higgs boson.  With over $30\xfb^{-1}$ it
may be possible to observe $h_{\rm inv}$ at the $3\sigma$ level
if its mass is below $\sim 125\gev$~\cite{Martin:1999qf}.
At the LHC, analyses  
indicate~\cite{Frederiksen:1994me}
that $m_{\hinv}$ may be probed up to $\sim 200\gev$ with $100\xfb^{-1}$.
It would be worthwhile to redo the
LHC analyses to take into account our current knowledge of SM particle
properties (parton distributions, top quark mass, etc.) and 
the current expectations for detector parameters, such as particle 
identification, tagging and energy resolution.

An NLC analyses of the invisibly~\cite{Eboli:1994bm} 
decaying Higgs boson indicates that
it can be probed very close to the kinematic limit of $\sqrt{s}-m_Z$.
Again, this is the general expectation of the $e^+e^-$ collider with
a beam constraint to search for peaks in the missing invariant mass
spectrum.  We expect that shared branching fractions into invisible states and 
SM states will be measured effectively at the NLC as well.  Nevertheless,
we encourage a detailed study on this important discovery issue, and
think that a comparison between NLC and LHC for
invisibly decaying Higgs boson
searches is more appropriate than comparing capabilities for discovery
of very large invariant mass resonances of strongly coupled EWSB.

\section{Impact on future experiment: seeing through the many ideas}

In summary, we have argued that a broad view of possible beyond-the-SM
theories combined with the accumulated data of ten years at LEP and SLC
indicate that we should expect a Higgs boson with mass 
less than $500\gev$.  This presents many important
discovery issues.  The most notable of these issues is how to discover
an invisibly decaying Higgs boson in this mass range.  Other issues 
arise if the Higgs boson mass turns out to be at the upper end of
this allowed range, having conspired with other ``new physics'' contributions
to satisfy the current EW data.  In both cases studied here,
top seesaw model with an extra quark and large extra dimensions with
KK gauge bosons at several TeV, the states are best resolved
by precision measurements at the NLC running on the $Z$ pole and
at higher energy, $\sqrt{s}=600\gev$.  Combining the results from
these measurements it is possible to observe the heavier scalar
and either the associated heavy quark or KK excitations of the gauge
bosons.  We think this result is probably general.

Finally, we emphasize what should be an obvious point to most: nothing
is metaphysically certain.  Certainty about what we may
or may not find at the next collider has {\it never} been a part
of the high-energy physics frontier.  Our results are not theorems,
but they are robust indications about what to bet on if one wants to
pursue the most likely directions for progress in our field.
If we knew what we were 
going to find there
would be no reason to build colliders.  Nevertheless, we think that
the last decade of experimental physics is paying off and is providing
us important clues that an NLC running at $\sqrt{s}\lsim 600\gev$
will be rewarding.  
The NLC will vastly improve our chances of finding the origin of
EWSB, and then would enable extraordinary precision EWSB measurements.

\bigskip
\noindent


\end{document}